\documentstyle[psfig,times]{mn}

\newif\ifAMStwofonts

\ifoldfss
  \ifCUPmtlplainloaded \else
    \NewTextAlphabet{textbfit} {cmbxti10} {}
    \NewTextAlphabet{textbfss} {cmssbx10} {}
    \NewMathAlphabet{mathbfit} {cmbxti10} {} 
    \NewMathAlphabet{mathbfss} {cmssbx10} {} 
  \fi
  \ifAMStwofonts
    \ifCUPmtlplainloaded \else
      \NewSymbolFont{upmath} {eurm10}
      \NewSymbolFont{AMSa} {msam10}
      \NewMathSymbol{\upi}     {0}{upmath}{19}
      \NewMathSymbol{\umu}     {0}{upmath}{16}
      \NewMathSymbol{\upartial}{0}{upmath}{40}
      \NewMathSymbol{\leqslant}{3}{AMSa}{36}
      \NewMathSymbol{\geqslant}{3}{AMSa}{3E}

    \fi
  \fi
\fi 

\ifnfssone
  \newmathalphabet{\mathit}
  \addtoversion{normal}{\mathit}{cmr}{m}{it}
  \addtoversion{bold}{\mathit}{cmr}{bx}{it}
  \newmathalphabet{\mathbfit} 
  \addtoversion{normal}{\mathbfit}{cmr}{bx}{it}
  \addtoversion{bold}{\mathbfit}{cmr}{bx}{it}
  \newmathalphabet{\mathbfss} 
  \addtoversion{normal}{\mathbfss}{cmss}{bx}{n}
  \addtoversion{bold}{\mathbfss}{cmss}{bx}{n}
  \ifAMStwofonts
    \ifCUPmtlplainloaded \else
      \UseAMStwoboldmath
      \makeatletter
      \new@mathgroup\upmath@group
      \define@mathgroup\mv@normal\upmath@group{eur}{m}{n}
      \define@mathgroup\mv@bold\upmath@group{eur}{b}{n}
      \edef\UPM{\hexnumber\upmath@group}
      \new@mathgroup\amsa@group
      \define@mathgroup\mv@normal\amsa@group{msa}{m}{n}
      \define@mathgroup\mv@bold\amsa@group{msa}{m}{n}
      \edef\AMSa{\hexnumber\amsa@group}
      \makeatother
      \mathchardef\upi="0\UPM19
      \mathchardef\umu="0\UPM16
      \mathchardef\upartial="0\UPM40
      \mathchardef\leqslant="3\AMSa36
      \mathchardef\geqslant="3\AMSa3E
    \fi
  \fi
\fi 

\ifnfsstwo
  \DeclareMathAlphabet{\mathbfit}{OT1}{cmr}{bx}{it}
  \SetMathAlphabet\mathbfit{bold}{OT1}{cmr}{bx}{it}
  \DeclareMathAlphabet{\mathbfss}{OT1}{cmss}{bx}{n}
  \SetMathAlphabet\mathbfss{bold}{OT1}{cmss}{bx}{n}
  \ifAMStwofonts
    \ifCUPmtlplainloaded \else
      \DeclareSymbolFont{UPM}{U}{eur}{m}{n}
      \SetSymbolFont{UPM}{bold}{U}{eur}{b}{n}
      \DeclareSymbolFont{AMSa}{U}{msa}{m}{n}
      \DeclareMathSymbol{\upi}{0}{UPM}{"19}
      \DeclareMathSymbol{\umu}{0}{UPM}{"16}
      \DeclareMathSymbol{\upartial}{0}{UPM}{"40}
      \DeclareMathSymbol{\leqslant}{3}{AMSa}{"36}
      \DeclareMathSymbol{\geqslant}{3}{AMSa}{"3E}
    \fi
  \fi
\fi 

\ifCUPmtlplainloaded \else
  \ifAMStwofonts \else 
    \def\upi{\pi}
    \def\umu{\mu}
    \def\upartial{\partial}
  \fi
\fi

\title{Can Supermassive Black Holes alter Cold Dark Matter cusps
through accretion?}
\author[J. I. Read and G. Gilmore]
       {J. I. Read and G. Gilmore \\
        Institute of Astronomy, Cambridge University, Madingley Road, Cambridge, CB3 0HA}
\date{Accepted.
      Received;
      in original form}

\pagerange{\pageref{firstpage}--\pageref{lastpage}}
\pubyear{2002}

\begin{document}

\maketitle

\label{firstpage}

\begin{abstract}
We present some simple models to determine whether or not the accretion
of cold dark matter by supermassive black holes is astrophysically
important. Contrary to some claims in the literature, we show that
supermassive black holes cannot significantly alter a power law
density cusp via accretion, whether during mergers or in the steady
state.

\end{abstract}


\section{Introduction}
Over the past few decades, the idea of dark matter has become deeply
rooted within the astrophysical community. In particular the
$\Lambda$CDM (cold dark matter with a cosmological constant) theory
has done very well in reproducing experimental results on large scales
(see e.g. Bahcall et. al. 1999 or Sellwood \& Kosowsky 2000).

However, on small scales (galaxies and smaller), the success of cold
dark matter has been less striking \cite{moore}. On galactic scales,
current experiments which measure the velocity dispersion of a galaxy
as a function of radius, have found enormous mass to light ratios in
all galaxies from dwarfs to giants \cite{salucci}. While this is
strong evidence for the existence of dark matter, the derived mass
profiles do not compare well with those predicted from simulations
involving cold dark matter \cite{moore}. In particular, cold dark
matter simulations predict the presence of a central cusp in the
density profile, whereas the observed profiles seem to have a well
defined core \cite{salucci}. This apparent misalignment of theory and
experiment has lead to three main types of solution: \\\\ 1. Dark
matter is real but not cold \cite{moore}.\\\\ 2. Dark matter is not
the solution and we should look to alternative theories
\cite{milgrom}.\\\\ 3. Cold dark matter is correct but current
theories are missing some important physics.\\

In other words: does the absence of a central dark matter cusp tell us
something about the nature of dark matter, or does it tell us
something about galaxy formation? Understanding this is essential for
facilitating direct search experiments for dark matter such as DAMA
\cite{dama}. With any number of particle candidates for dark
matter and the difficulty of removing noise from direct detection
experiments, particle physicists almost need to know what they are
looking for before they can find it. As such, any complimentary method
which can rule out some candidates is extremely
important. Unfortunately, before any robust conclusions about the
nature of dark matter can be made from the rotation curve measurements
of nearby galaxies, we need to be sure that what we are measuring is
the dark matter profile as it would have been before the formation of
a galaxy. It is this profile which contains information about the
detailed nature of dark matter. If the profile has been changed due to
astrophysical effects such as matter outflow \cite{gnedin} or bars
\cite{athanassoula} then some or maybe all of the information about
the nature of dark matter could have been lost. It is
important to work out which astrophysical effects we should consider
and which are not important.   

In this paper, we explore the fate of a CDM cusp in the presence of
a central black hole. Recently, Zhao, Haehnelt \& Rees 2001
(hereafter ZHR) have suggested that tidal stirring from infalling
satellites could lead to significant dark matter accretion, accounting
for some 20-40\% of the mass of a central supermassive black hole if
loss cone refilling can be efficient enough. As
such, we consider whether or not supermassive black holes can alter
the dark matter distribution in a galaxy in any significant way.

In section 2, we recall that the steady state accretion rate of cold
dark matter onto a supermassive black hole is negligible. In section
3, we present a toy model for calculating the mass of cold dark matter
accreted during merger events. In section 4, we discuss the validity
of our model and the subsequent effect on the underlying dark matter
distribution. Finally, in section 5 we conclude.

\section{The Steady State Accretion Rate}
The steady state accretion rate of matter onto a black hole has been
treated by several authors (e.g. Lightman \& Shapiro 1977). Any
particles which lie on orbits which bring them within the Schwarzschild radius
of the hole (see equation \ref{eqn:schwartz}) will be swallowed. Since
both low energy, high angular momentum and high energy, low angular
momentum particles can be swallowed the problem is necessarily
two-dimensional. The fraction of particles in phase space which will
be swallowed is described by the `loss cone' \cite{light} and any
particles lying within the loss cone will be swept away in a dynamical
time $\sim$ the time taken for a particle to complete one orbit.

\begin{figure}
\psfig{figure=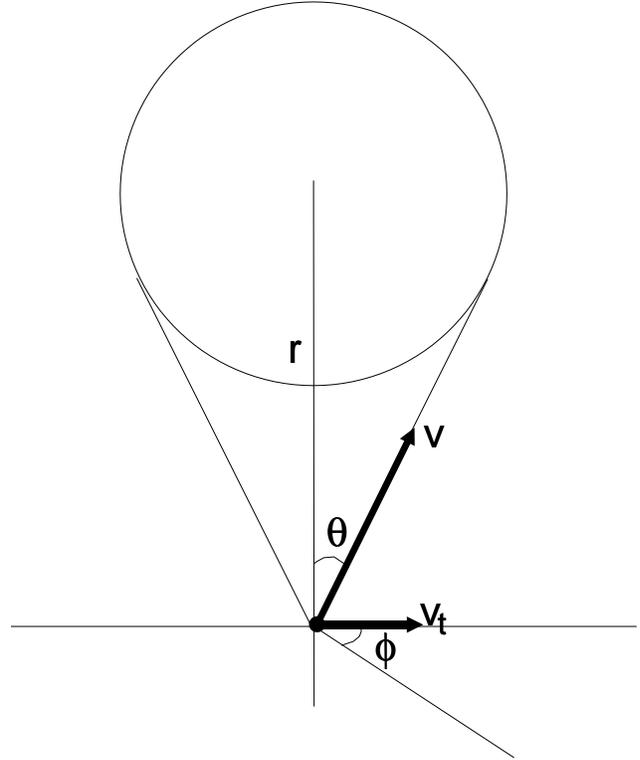,width=0.5\textwidth,angle=0}
\caption{Schematic diagram of a black hole and its loss cone. The
solid circle at the origin represents a dark matter particle a
distance r away from a black hole (depicted by the large circle). The
dark matter particle has velocity, $v$ and transverse velocity $v_t =
L_{max}/r$.}
\label{fig:losscone}
\end{figure}

As can be seen in figure \ref{fig:losscone}, the loss cone can be
thought of as the locus of points swept out from trying to point a
particle at the black hole from some distance r away.

Now, by considering particle orbits in a Schwarzschild metric
\cite{gr}, any particle with an angular momentum per unit mass less
than $L_{max} = \sqrt{12}GM_{BH}/c$ will fall within the
Schwarzschild radius and be swallowed (where $M_{BH}$ is the mass of
the black hole, c is the speed of light in vacuo and G is the
gravitational constant).

The solid angle subtended by the loss cone, $\Omega$ can be written
as:

\begin{eqnarray}
\Omega & = & \int_{0}^{2\pi}\int_{0}^{\theta}{\sin{\theta'}}d\theta'
d\phi \nonumber \\ & \simeq & \pi{\theta}^2 \simeq \pi({v_t}/v)^2 =
\pi({L_{max}}/rv)^2
\label{eqn:omega}
\end{eqnarray}

So, given some phase space density of dark matter particles, $f(r,v)$,
the total mass inside the loss cone $M_{lc}$ is given by:

\begin{eqnarray}
M_{lc} & = & \int_{0}^{\infty} \int_{r_s}^{r_{200}} {{\Omega}
4{\pi}r^2 4{\pi}v^2 f(r,v)} dr dv \nonumber \\ & = & \int_{0}^{\infty}
\int_{r_s}^{r_{200}} {{16\pi^3{L_{max}}^2}f(r,v)} dr dv
\label{eqn:losscone}
\end{eqnarray}

Notice that the lower limit for the integral over r is the
Schwarzschild radius. This is because the black hole itself must
provide a cut-off for the dark matter distribution. The upper limit is
given by $r_{200}$, the virial radius \cite{NFW}.

Assuming a Maxwell Boltzmann distribution for the velocity density of
the dark matter particles and an NFW profile \cite{NFW} for the
spatial density gives us:
 
\begin{eqnarray}
f(r,v) & = & \rho_{NFW}(r) g_{MB}(v) \nonumber \\ & = &
{\frac{\rho_0}{(\frac{r}{r_0})(1+\frac{r}{r_0})^2}}
\left(\frac{3}{2<v^2>\pi}\right)^{\frac{3}{2}}
e^{\left({\frac{-3v^2}{2<v^2>}}\right)}
\label{eqn:maxwell}
\end{eqnarray}

Where $\rho_0$ and $r_0$ are the scale factors for a particular
galaxy, and $<v^2>^{1/2}$ is the velocity dispersion of the dark
matter halo. The velocity dispersion is taken to be a constant, which
is a reasonable approximation given the observed rotation curves in
many galaxies.

For a $10^6$ M$_\odot$ black hole at the centre of a Milky Way type
galaxy (we take scaling values as in table \ref{tab:initial}), the dark
mass within the loss cone is $\sim$0.6 M$_\odot$, while the total mass
of dark matter is some $1.9 \times 10^{12}$M$_\odot$. Thus a tiny
fraction of the dark matter is accreted in a crossing time but then
not much else can happen until the loss cone is refilled. For stars
refilling can be achieved by two-body interactions, continuing satellite
mergers and dynamical friction, allowing a continual
diffusion of matter into the loss cone \cite{light}. However, for cold
dark matter, the interaction cross section is presumed to be so small
that the time taken to re-fill the loss cone by diffusion is very much
longer than the age of the universe.

The problem can be avoided by postulating that dark matter can
self-interact thus raising the interaction cross section and allowing
significant steady-state accretion rates \cite{ostriker}. ZHR have
also suggested that tidal stirring from infalling satellites could
refill the loss cone. We look at this idea in more detail in the
following section.

\section{Accretion during merger events}
As shown above, the steady state accretion rate of dark matter is
negligible. However, during a galaxy merger event, while a central
supermassive black hole is stationary with respect to the dark
matter in its host galaxy, it is not stationary with respect to the dark
matter in the merger galaxy. In the rest frame of one of the galaxies,
the central black hole, belonging to the other galaxy will spiral in,
sweeping up dark matter on its way. As the two galaxies merge, their
CDM halos will be tidally stripped out to the Roche limit \cite{roche}
where the density of a halo approximately matches that of its
surroundings. We can expect, then, that the CDM cusps from the two
galaxies should make it right into the centre of the forming merger remnant.

This idea is borne out by various numerical studies (see
e.g. Van Albada \& Van Gorkom 1977 \& more recently Milosavljevic \&
Merritt 2002) which show that the initial conditions are statistically
preserved in major merger events. As such, it is not clear that a dark
matter cusp will survive during galaxy mergers. Two black holes could,
in principle, mutually accrete dark matter from each other's host
cusp. The magnitude of the effect will depend on the loss cone
refilling rate during such a merger process and it is this that we
will discuss in 3.1 and 3.2.

The presence of a significant bulge component in galaxies, apparently
correlated with the central black hole mass, can set a limit on the
efficiency of this process. If the cusp depletion is found to be very
efficient then it must occur in the early stages of bulge growth, or
in low surface brightness (LSB) galaxies where the dark matter
dominates over the baryonic matter even in the central regions.

Faber et. al 1997 examined the central density profile of 61
elliptical galaxies and spiral bulges. They found that, while all of
the galaxies in their sample were well fit by a power law profile, the
brightest galaxies contained a central core in the light profile at a break
radius of $\sim 10-100$ parsecs. They suggest, as have others
(e.g. Quinlan 1996 and  Milosavljevic \& Merritt 2001, hereafter
MM), that the central cusp depletion could be due to scattering
during major merger events as two massive black holes form a hard
binary. We consider a complimentary mechanism, operating earlier in the
merger process, based on the accretion of scattered material during
galaxy mergers. We show that this mechanism is not important for
growing black holes or rearranging dark matter cusps.

Although in our analysis we will talk mainly about the effect
on the central dark matter, our analysis could be applied equally to
the central baryonic matter cusps which are observed in bright galaxies.

We treat the infalling black hole as a scattering body which spirals
into the centre of a galaxy via dynamical friction. Some of the dark
matter which it scatters will be scattered back onto itself - {\it
self scattering}, and some will be scattered into the loss cone of the
other hole - {\it mutual scattering}. The sum of these two effects
will give us an estimate of the total accretion rate of dark
matter. We now look at these effects quantitatively.

\subsection{Self Scattering}
Consider two in-spiralling black holes, A and B, from the reference
frame stationary relative to hole B. In the {\it self scattering} case
we consider the in-spiralling black hole, A, as an effective cross
sectional area sweeping through phase space.

We need only consider the growth of hole A though, by symmetry, hole B
will also grow via this mechanism. We can then parameterise hole B
solely by the nature of its host galaxy. This is because of the
empirically observed correlation between black hole mass and bulge
velocity dispersion \cite{ferrarese}. Small black holes reside in
shallow potential wells, larger black holes reside in larger potential
wells.

We assume that hole A is brought in via dynamical friction from the
dark matter and baryonic matter in the centre of the stationary galaxy
(hole B's host galaxy). Following the Chandrasekhar prescription
(c.f. Binney \& Tremaine 1987), we have that the inward velocity of
the hole, $\frac{dr}{dt}$ is given by:

\begin{equation}
\frac{dr}{dt} = \frac{-KM_A}{v_c^3}r\rho(r)
\label{eqn:dynf}
\end{equation}

where, $\rho(r)$ is the density distribution of the central matter
(baryons and dark matter), $M_A$ is the mass of hole A, $v_c(r)$ is
the local circular velocity and K is approximately a constant. For a
Maxwell Boltzmann distribution of particle velocities, $K$ is given by
\cite{binney}:

\begin{eqnarray}
K & = & 0.428G^22\pi \ln(1+\Lambda^2) \\ \Lambda & = &
\frac{b_{max}<v^2>}{G M_A}
\end{eqnarray}

where, $G$ is the gravitational constant, $b_{max}$ is the maximum
impact parameter for the in-spiralling hole and
${<v^2>}^{\frac{1}{2}}$ is the mean velocity dispersion of the
surrounding baryonic and dark matter.
  
From equation \ref{eqn:dynf}, we can derive the dynamical friction
infall time $t_{dyn}$ for the hole starting at radius $r_i$. This
gives us:

\begin{equation}
t_{dyn} = \int_0^{r_i}\frac{1}{KM_A}{\frac{v_c(r)^3}{r\rho(r)}} dr
\label{eqn:tdynf}
\end{equation}

Equation \ref{eqn:tdynf} is the Chandrasekhar equation for dynamical
friction and is approximately valid provided that the infalling mass
does not exceed the mass interior to its orbit \cite{binney}. Notice
that the minus sign in equation \ref{eqn:dynf} has been substituted for
swapped integration limits.

We can also use equation \ref{eqn:dynf} to derive the amount of mass
swept up by the black hole as it moves from a radius $a$ to a radius
$b$. The distance travelled by the hole in a time $dt$ is given by:

\begin{equation}
ds = dtv_c(r) = \frac{v_c(r)^4dr}{Kr\rho(r)}
\end{equation}

If the black hole cross section is given by $\sigma_{A}(r)$ then the
mass of dark matter swept up by the hole by moving through a distance
$ds$ is then:

\begin{equation}
dM_A = \frac{v_c(r)^4\sigma_{A}(r)\rho(r)dr}{Kr\rho(r)}
\end{equation}

and so the total mass swept up as the hole moves from a radius $a$ to
$b$ is:

\begin{equation}
\Delta M_A = \int_{b}^{a} {dM_A} = 
\int_{b}^{a} {\frac{1}{KM_A}\frac{v_c(r)^4\sigma_{A}(r)}{r}}dr
\label{eqn:mdynf}
\end{equation}

\subsubsection{The black hole cross section}
If the black hole cross section were just $\pi r_s^2$, where $r_s$ is
the Schwarzschild radius, then the volume swept up in phase space
would be truly tiny. The Schwarzschild radius is given by:

\begin{equation}
r_s=\frac{2GM_A}{c^2}
\label{eqn:schwartz}
\end{equation}

where $c$ is the speed of light in vacuo. So, for a $10^6 M_\odot$
black hole, $r_s \sim 10^{-7}$ parsecs.

However, a massive body such as a black hole interacts with the medium
through which it moves, preferentially scattering particles onto
orbits which will bring them within the Schwarzschild radius,
allowing them to be swallowed. Taking this {\it gravitational
focusing} into account gives us a total cross section of
\cite{binney}:

\begin{eqnarray}
\sigma_{A}(r) & = & \pi r_s^2 + \frac{2\pi GM_{A}r_s}{v_c(r)^2}
\nonumber \\ & = & \pi r_s^2 \left(1+\frac{c^2}{v_c(r)^2}\right)
\nonumber \\ & \simeq & \frac{4\pi G^2M_A^2}{c^2v_c(r)^2}
\end{eqnarray}

Since $v_c(r)$ must always be less than the speed of light, the cross
section due to gravitational focusing will always be much more
important than the physical cross section of the black hole.

Putting this all together we get, for the mass swallowed as the hole
moves from radius $a$ to $b$:

\begin{eqnarray}
\Delta M_A & = & \int_{b}^{a} {AM_A\frac{v_c(r)^2}{r}}dr
\label{eqn:madynf} \\
A & = & \frac{2}{0.428\ln(1+\Lambda^2)c^2}
\label{eqn:A}
\end{eqnarray}

Using equation \ref{eqn:madynf} we can now calculate the mass swept up
as the black hole spirals in. We cannot, however, naively integrate
from some start radius to zero to obtain the mass swallowed. This is
because, if the black hole manages to complete a whole orbit before
moving inwards a distance greater than $\sim \sqrt{\sigma_{A}}$ then
we have the problem of shell crossing; that is, we would be counting
the mass more than once. In order to avoid this problem, we can make
the distance $a$ to $b$ some small interval and then numerically sum
each mass shell $\Delta m$, comparing each $\Delta m$ with the actual
mass of cold dark matter present, and taking the smaller value. In
this way, we avoid divergent mass consumption at small radii where the
same part of configuration space is swept out many times but the mass within
that space can be swept up only once. Quantitatively, this means that
we require:

\begin{equation}
\Delta M_A < 4\pi\int_{b}^{a}{r^2\rho(r)}dr \hspace{0.1in}\forall a,b
\end{equation}

\begin{figure}
\psfig{figure=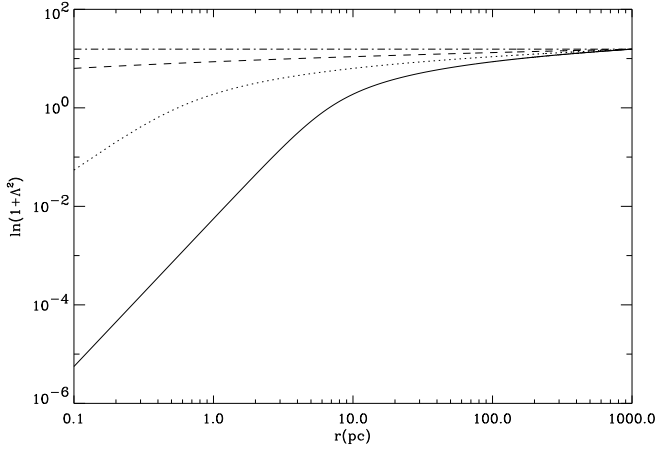,width=0.5\textwidth,angle=0}
\caption{Radial dependence of the Coulomb factor, $\ln(1+\Lambda^2)$,
in power law density profiles with exponent $\alpha$. The solid line
is for $\alpha=0.5$, the dotted line is for $\alpha=1$, the dashed
line is for $\alpha=1.5$ and the dot-dashed line is for $\alpha=2$.}
\label{fig:lambda}
\end{figure}

\subsubsection{The Density Profile}

The density profile for the dark and luminous matter in most galaxies
can be well approximated, within a few kiloparsecs of the centre, by a
power law \cite{blais}. As such, we adopt the following form for $\rho(r)$:

\begin{equation}
\rho(r) = \rho_0\left(\frac{r}{r_0}\right)^{-\alpha}
\label{eqn:power}
\end{equation}

For the dark matter, $\alpha$ is observed to be typically $\alpha<1$,
whereas CDM simulations predict dark matter slopes of $\alpha>1$
\cite{blais}. 

The local circular velocity is then given by \cite{binney}:

\begin{eqnarray}
v_c(r)^2 & = & \frac{4\pi G}{r} \int_0^r {r'^2\rho(r')} dr' \nonumber
\\ & = & \frac{4\pi G\rho_0r_0^\alpha}{3-\alpha}r^{2-\alpha}
\label{eqn:initvel}
\end{eqnarray}

For a power law profile, it can be shown (see Appendix I) that:

\begin{equation}
<v^2> \simeq v_c^2
\label{eqn:apenI}
\end{equation}

and if the maximum impact parameter, $b_{max}$, is taken to be the
starting radius of the hole, $r_i$, this gives us:

\begin{equation}
\Lambda = \frac{4\pi\rho_0r_0^\alpha r_i}{M_A(3-\alpha)}r^{2-\alpha}
\end{equation}

Figure \ref{fig:lambda} plots $\ln(1+\Lambda^2)$ against $r$ for
initial conditions set up as in table \ref{tab:initial} and for
varying $\alpha$. For $r > 0.01$kpc, or $\alpha > 0.5$,
$\ln(1+\Lambda^2) \sim$ constant. However, in general it is not
constant over the full range of $r$ and hence neither is $A$ (see
equation \ref{eqn:A}). We assume that $A$ is constant over the small
change in $r$, $\Delta r = a-b$, but does vary as a function of $a$.

Thus, putting this all together gives us, for the mass swept up by
hole A:

\begin{eqnarray}
\Delta M_A & = & \frac{4\pi
GA(a)M_A\rho_0r_0^\alpha}{(3-\alpha)(2-\alpha)}(a^{2-\alpha} -
b^{2-\alpha}) \hspace{0.1in} \alpha<3,\alpha\neq 2 \nonumber \\ \Delta
M_A & = & 4\pi GA(a)M_A\rho_0r_0^2\ln(\frac{a}{b})
\hspace{1in} \alpha=2 \nonumber \\ &&
\label{eqn:deltama}
\end{eqnarray}

\begin{table}
\psfig{figure=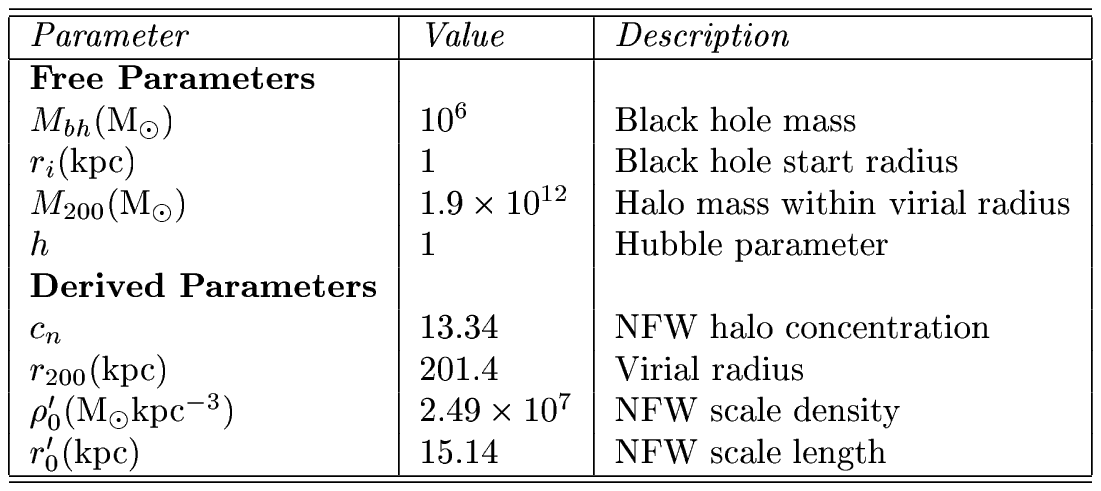,width=0.5\textwidth,angle=0}
\caption{Initial Conditions. See text for details.}
\label{tab:initial}
\end{table} 

\subsubsection{Initial Conditions}
From equation \ref{eqn:deltama}, we can see that the important factors
for determining the amount of mass accreted are $M_A$, the mass of the
hole, and $\rho_0$, $r_0$ and $\alpha$, the density parameters for the
central cusp.

The initial infall radius of the hole can be taken from numerical
simulations performed by MM. We use 1kpc as our starting position for
the hole.

The density parameters are scaled such that the galaxy has the same
mass as an NFW-type profile within 1kpc. The NFW profile is given by
\cite{NFW}:

\begin{eqnarray}
\rho_{NFW} & = & \frac{\rho_0'}{(\frac{r}{r_0'})(1+\frac{r}{r_0'})^2}
\\ \lim_{r<<r_0'} \rho_{NFW} & \simeq & \frac{\rho_0' r_0'}{r}
\label{eqn:dark}
\end{eqnarray}

Where $\rho_0'$ and $r_0'$ are given by:

\begin{eqnarray}
\rho_0' & = & \delta_0\rho_{crit} \\ \delta_0 & = &
\frac{200}{3}\frac{c_n^3}{\ln(1+c_n)-\frac{c_n}{1+c_n}} \\ r_0' & = &
r_{200}/c_n \\ r_{200} & = & \left(\frac{M_{200}}{\frac{4}{3}\pi
200\rho_{crit}}\right)^{\frac{1}{3}}
\end{eqnarray}

Where $r_{200}$ is the virial radius, $M_{200}$ is the mass within the
virial radius and is taken to be the mass of the galaxy, $c_n$ is the
NFW concentration parameter, and $\rho_{crit} = 277.3h^2$M$_\odot$
kpc$^{-3}$ is the critical density.

The only free parameter in the NFW profile is $M_{200}$ since the
concentration parameter is found, numerically, to be a function of the
mass of a galaxy \cite{NFW}.

The power law scale density and radius ($\rho_0$, $r_0$) we use are
then scaled to contain the same mass as the NFW profile within
1kpc. Thus:

\begin{eqnarray}
r_0 & = & r_0' \\ \rho_0 & = &
\rho_0'\left(\frac{r_0}{1kpc}\right)^{1-\alpha}\frac{(3-\alpha)}{2}
\end{eqnarray}

Recall that this density profile describes the galaxy which is in the
stationary frame. The infalling galaxy is parameterised only by the
mass of its black hole. Thus, by setting up the profiles for a Milky
Way type galaxy, we can model the full range of mergers, from equal
mass to minor mergers, by scaling the infalling black hole mass.

A summary of all of the initial conditions used is given in table
\ref{tab:initial}. The parameters are broken up into those which are
put into the model and those which are subsequently derived. The
scaling parameters for the NFW profile, for example, are set up
consistently for a halo which contains a mass of $1.9 \times
10^{12}$M$_\odot$ within the virial radius. A halo of this mass has a
concentration parameter of about 13.3 \cite{NFW}.

\subsubsection{Results}
\begin{figure}
\psfig{figure=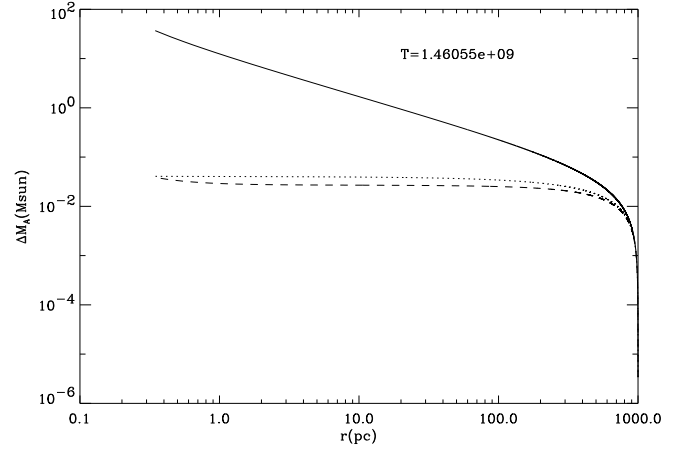,width=0.5\textwidth,angle=0}
\caption{Mass accreted by a $10^6$M$_\odot$ black hole spiralling in
to a Milky Way type galaxy from 1 kiloparsec away from the centre due
to self scattering. The infall time, T, is marked in years. The solid
curve is for $\alpha=2.9$, the dotted curve is for $\alpha=1$ and the
dashed curve is for $\alpha=0.5$.}
\label{fig:selfscat}
\end{figure}

The numerical integration was performed using IDL, with the initial
conditions set up as in table \ref{tab:initial} and with a numerical
resolution of $\Delta r = a-b = 0.1$ parsecs. The results are
displayed in figure \ref{fig:selfscat}.

As can be seen, most of the accretion occurs in the outer regions and
increases with $\alpha$. However, even for $\alpha=2.9$, the effect is
small, with less than $100$M$_\odot$ accreted in total. This is a tiny
fraction of the black hole's mass, which is some $10^6$M$_\odot$.  For
an NFW profile, with $\alpha=1$ (dotted line, figure
\ref{fig:selfscat}), the total mass accreted is less than in the
steady state case. Finally, from equation \ref{eqn:deltama}, we can
see that the total mass accreted scales linearly with $M_A$, so the
fractional increase in black hole mass is unaffected by any increase
in $M_A$.

\subsection{Mutual Scattering}
Now we consider the case where the black holes scatter dark matter
into each others' loss cone. Consider, as in the self scattering
scenario, a black hole, A, spiralling into a hole B, from a reference
frame stationary with respect to the hole B. As before, each black
hole has its own bound CDM cusp. The dark matter hole A scatters, for
example, is that which resides in hole B's CDM cusp.

Hole A will scatter some fraction, $\eta_A$, of particles within its
sphere of influence into the loss cone of hole B, and these particles
will then be quickly swallowed in a dynamical time. Similarly, hole B
will scatter particles into hole A's loss cone and hence both holes
will grow via accretion as they merge.

The problem, then, is similar to the self scattering case except that
now, the effective black hole cross section is due to the sphere of
influence of hole A, and hole A causes the growth of hole B.

Since each hole induces the growth of the other, we should consider
both $M_A$ and $M_B$ as functions of $r$ and consider the growth of
both holes. However, we will make the assumption that $\frac{\Delta
M_A}{M_A} \sim \frac{\Delta M_B}{M_B} \sim$ small so that $M_A$ and
$M_B$ are $\sim$ constant.

The sphere of influence of hole A is given by:

\begin{equation}
r_{sph,a} = \frac{GM_A}{<v^2>}
\label{eqn:sphinf}
\end{equation}

Where $M_A$ is the mass of hole A and ${<v^2>}^{1/2}$ is the local
spherical velocity dispersion in hole B's host galaxy. For a
$10^6$M$_\odot$ black hole in a Milky Way type galaxy, $r_{sph,a} \sim
0.3$ parsecs. Thus the effective black hole cross sectional area, $\pi
r_{sph,a}^2$ is $\sim 0.1 $pc$^2$, whereas the effective cross
sectional area for the self scattering case was just $\sim 10^{-6}$
pc$^2$. As such, if the scattering efficiency, $\eta_A$,is large
enough, this could be an important mechanism for growing black holes.

From equations \ref{eqn:mdynf} and \ref{eqn:sphinf}, we get for the
mass accreted by hole B as black hole A moves from a radius $a$ to
$b$:

\begin{equation}
\Delta M_B = \int_{b}^{a}{\frac{\eta_A\pi
G^2M_A}{K}\frac{v_c(r)^4}{<v^2>^2r}}dr
\end{equation}

For a power law density profile (see equation \ref{eqn:power}) and
using equation \ref{eqn:apenI}, this gives us:

\begin{eqnarray}
\Delta M_B & = & A'(a)M_A\eta_A\ln(\frac{a}{b})
\label{eqn:deltamb} \\
A' & = & \frac{1}{0.856\ln(1+\Lambda^2)}
\label{eqn:A'}
\end{eqnarray}

\subsubsection{The scattering efficiency}
Comparing equations \ref{eqn:deltama} and \ref{eqn:deltamb}, we can
see straight away that, if the scattering efficiency, $\eta_A$, were
nearly 1 then nearly all of the central dark matter would be swept up
as the holes spiral together. This is because the constant, $A'$ is
larger than $A$ by a factor $\sim c^2$. As such, the value of $\eta_A$
is key to determining the magnitude of the accretion effect.

We can obtain a crude estimate for $\eta_A$ if we assume that all dark
matter particles are scattered randomly. If this is the case then the
probability, $P_A$, that a particle will be scattered into hole B's
loss cone will be simply the fraction of phase space subtended by its
loss cone.

Thus we get, from equations \ref{eqn:omega}, \ref{eqn:maxwell},
\ref{eqn:initvel} and \ref{eqn:apenI}:

\begin{eqnarray}
P_A & \simeq & \frac{1}{4\pi}\int_{0}^{\infty}{\Omega4\pi v^2g(v)}dv
\\ & = & \frac{9GM_B^2(3-\alpha)}{2\pi c^2\rho_0r_0^\alpha
r^{(4-\alpha)}}
\label{eqn:P_A}
\end{eqnarray}

If $N_A$ is the number of times hole A crosses the same section of
configuration space, then the fraction of particles scattered into B's
loss cone will be: 

\begin{equation}
\eta_A = 1-(1-P_A)^{N_A}
\label{eqn:eta}
\end{equation}

$N_A$ is essentially the number of attempts hole A can make at
scattering the same particles into hole B's loss cone.

Thus, if the time taken for hole A to complete one orbit at a radius,
$r$ is given by $T(r) = \frac{2\pi r}{v_c(r)}$, then $N_A$ is given
by:

\begin{eqnarray}
dN_A & \simeq & \frac{dt}{T(r)} \nonumber \\ & = &
\frac{v_c(r)^4}{2\pi KM_Ar^2\rho(r)}dr \nonumber \\ \Rightarrow & &
\nonumber \\ N_A & = &
\frac{8A'(a)\rho_0r_0^\alpha}{M_A(3-\alpha)^3}(a^{3-\alpha}-b^{3-\alpha})
\label{eqn:N_A}
\end{eqnarray}

So that $N_A$ is now the number of times hole A crosses the same
section of phase space as it moves from a radius $a$ to a radius $b$.
Thus, if $a-b=\Delta r \ll a$, such that $P_A(a) \simeq P_A(b)$ and if
$\eta_A \ll 1$ then:

\begin{eqnarray}
\eta_A(a) & \simeq & 1-e^{-N_AP_A} \nonumber \\
& \simeq & 1-\exp\left(\frac{-36GA'(a)M_B^2(a-b)}{\pi c^2M_A(3-\alpha)a^2}\right)
\label{eqn:aproxeta}
\end{eqnarray}

Notice that the scattering efficiency is maximised for $M_A \ll
M_B$. This is because the size of hole B's loss cone is proportional
to $M_B^2$ (see equation \ref{eqn:losscone}) and so as $M_B$
increases, the chance of hole A scattering matter into B's loss cone
increases as $M_B^2$. As $M_A$ increases, the infall time for hole A
decreases and so hole A will cross the same section of configuration
space fewer times, reducing the scattering efficiency.

The dependence of the scattering efficiency on $\alpha$ is seemingly
very weak. However, $A'$ is a strong function of $\alpha$,
particularly for small $r$ (see figure \ref{fig:lambda} and equation
\ref{eqn:A'}). Since $A'$ increases as $\alpha$ is reduced, we can
expect that $\eta_A$ will be similarly correlated.

Notice also that $\eta_A$ is independent of $\rho_0$ and $r_0$. This
is because increasing the central density normalisation
$\rho_0r_0^\alpha$, while keeping the black hole mass constant reduces
the fraction of phase space subtended by the loss cone, and so $P_A$
falls. However, the number of times hole A sweeps through the same
section of phase space, $N_A$, is subsequently increased. The two
effects cancel out, so that the scattering efficiency depends only on
the slope of the density profile and not its normalisation. 

\begin{figure}
\psfig{figure=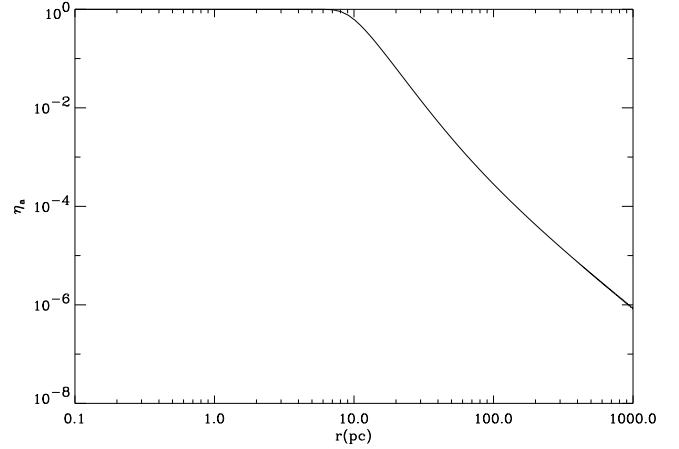,width=0.5\textwidth,angle=0}
\caption{The scattering efficiency $\eta_A$, as a function of radius
for $M_A=M_B=10^8M_\odot$, $\alpha = 1$.}
\label{fig:scateff}
\end{figure}

Finally, notice that the scattering efficiency is a strong function of
radius and will be maximised for small $a$. For most radii, the
scattering efficiency is absolutely tiny. Figure \ref{fig:scateff}
plots $\eta_A(a)$ against $a$ for $M_A=M_B=10^8M_\odot$, $\alpha =
1$. As can be seen for almost all radii, as might be expected
physically, the scattering efficiency is practically zero. Only when
the holes are close does it become large enough to produce
any significant amount of accretion. 

\subsubsection{Results}
\begin{figure}
\psfig{figure=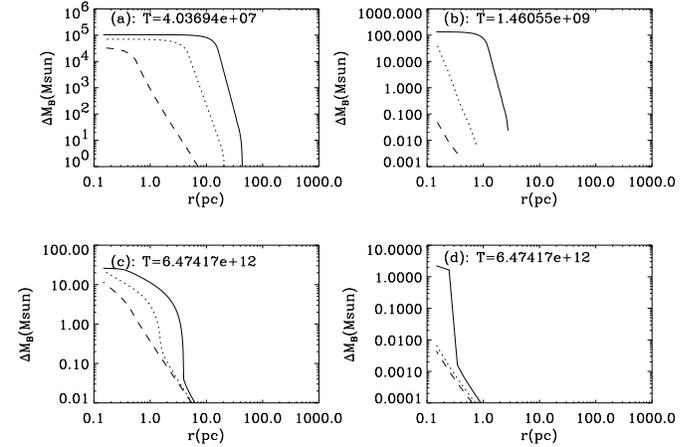,width=0.5\textwidth,angle=0}
\caption{Mass accreted by hole B due to the scattering of dark matter
particles by hole A. Graph (a) is for $M_A=M_B=10^8$M$_\odot$. Graph
(b) is for $M_A=M_B=10^6$M$_\odot$. Graph (c) is for
$M_A=10^2$M$_\odot$, $M_B=10^8$M$_\odot$. Graph (d) is for
$M_A=10^2$M$_\odot$, $M_B=10^6$M$_\odot$. The solid lines are for
$\alpha=0.5$, the dotted lines are for $\alpha=1$ and the dashed lines
are for $\alpha=1.5$. The infall times, T, are marked in years for
each case.}
\label{fig:mutescat}
\end{figure} 

The numerical integration was performed using IDL, with the initial
conditions set up as in table \ref{tab:initial} and with a numerical
resolution of $\Delta r = a-b = 0.1$ parsecs. The results are
displayed in figure \ref{fig:mutescat}.

As can be seen from equations \ref{eqn:aproxeta},
\ref{eqn:deltamb} and figure \ref{fig:scateff}, accretion becomes
significant only for small $r$. Reducing $\alpha$, the slope of the
cusp, pushes the radius at which accretion becomes significant to
larger radii but does not significantly affect the total mass accreted.

\begin{figure}
\psfig{figure=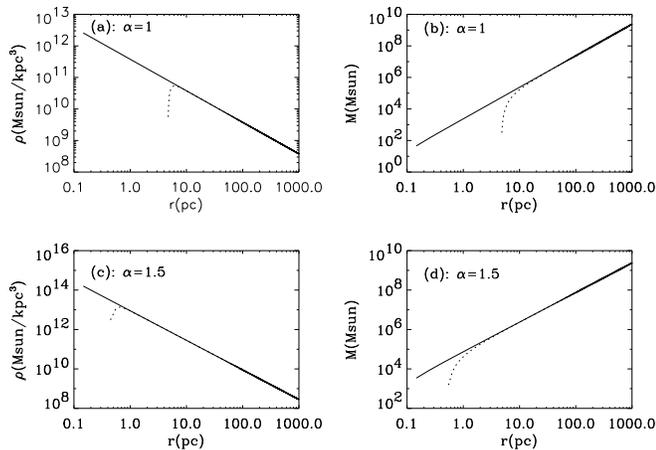,width=0.5\textwidth,angle=0}
\caption{The change in the central dark matter density due to
accretion onto two merging $10^8$M$_\odot$ black holes. Graphs (a) and
(c) show the central density before (solid line) and after (dotted
line) the merger, for an initial density cusp of slope, $\alpha=1$ and
$\alpha=1.5$ respectively. Graphs (b) and (d) show the central mass
interior to $r$ before (solid line) and after (dotted line) the
merger, also for an initial density cusp of slope, $\alpha=1$ and
$\alpha=1.5$ respectively.}
\label{fig:newden82}
\end{figure}

Figure 5(a) shows an extreme case - for $M_A=M_B=10^8$M$_\odot$. Here,
some $10^5$M$_\odot$ is accreted. While this amounts to only $\sim
0.1\%$ of the mass of hole B, it is sufficient to completely remove
all of the dark matter within $r_c \sim 20$ parsecs (see figure
\ref{fig:newden82}). Figure 5(b) shows a more likely scenario - for
$M_A=M_B=10^6$M$_\odot$. For an NFW type inner slope of $\alpha$=1 or
shallower, $\sim 100$M$_\odot$ is accreted. However for steeper slopes
this rapidly falls to a mere $0.1$M$_\odot$, which is accreted only
within the inner few parsecs. Figure 5(c) shows the case for
$M_A=100$M$_\odot$, $M_B=10^8$M$_\odot$. While only a few tens of
solar masses are accreted, in the early stages of black hole growth,
many low mass holes could have coalesced to form a supermassive
central black hole. Miralda-Escude \& Gould (2000) estimate that some
25,000 low mass black holes would have fallen into the central few
parsecs of the Milky Way by the present day. If each of these scatters
$\sim 10$M$_\odot$ of dark matter onto the central black hole then all
of the dark matter within $r_c \sim 0.01$ parsecs ($\sim
2.4\times10^5$M$_\odot$) would be swept up. As with the case (a), this
amounts to only $\sim 0.1\%$ of the mass of hole B. Finally, Figure 5(d)
shows the case for $M_A=100$M$_\odot$, $M_B=10^6$M$_\odot$. As with
the case (c), many infalling black holes could lead to all of the dark
matter within $r_c \sim 1$ parsec or so being swept up. However, this
amounts to only $\sim 2000$M$_\odot$, again only $\sim 0.1\%$ of the
mass of hole B.

\section{Discussion}
We have looked at some simple models which treat black holes in spherical
systems as scattering bodies. A fraction of the scattered material is
removed by accretion onto the scattering bodies, either through self
scattering or mutual scattering as discussed above.

Self scattering is most effective when the black holes are at their
maximum separation, and for steep central density slopes. However,
even in the most extreme situations, the overall mass accreted is
negligible, and usually less than that accreted in the steady state
scenario.

Mutual scattering is more effective, but only at very small radii. The
main problem with this is that at radii of a few to a few tens of
parsecs, the black holes are likely to form a stable hard binary
\cite{merritt}. Dynamical friction then no longer applies since, as the
binary hardens, much of the central matter is scattered away leaving
little to disrupt the binary system. This mechanism, proposed by
Quinlan (1996) and others, is effective on a scale of hundreds of
parsecs. However, dense central clusters are found around massive
black holes, so this process cannot be universal. Lauer et al. (2002)
have recently identified several galaxies where central luminosity
profiles are seen with cores of $\sim 20$ parsecs. It seems likely,
then, that well before any significant accretion could take place,
much of the central matter would have already been removed by the
formation of a hard supermassive black hole binary. This binary system
would then decay, in the usual way, via gravitational radiation.

The mutual scattering mechanism can, however, also be applied to
infalling massive satellites. A small bound system would survive right
into the centre but be disrupted well before the formation of a
binary. It seems unlikely, however, that the satellite would remain
bound as close as 10 parsecs to a supermassive black hole. The main
point though is that, {\it even if this were the case}, the total mass
accreted would still be a tiny fraction of the central black hole's
starting mass and would affect a cold dark matter cusp only on the
scale of a few parsecs. It seems that supermassive black holes
cannot alter a central dark matter distribution on the kiloparsec
scale required to align cold dark matter theory predictions with data
from rotation curves. 

Finally, removing the constraint of spherical symmetry is unlikely to
drastically alter our results. Although the model we present is rather
crude, the magnitude of the accretion affect is found to be so small,
that it is hard to imagine how a change in initial assumptions could
lead to significant accretion. While it is true that particles lying
on highly eccentric orbits are more likely to be accreted, the effect
of strong dynamical friction as required in our model is likely to
circularise the orbits of the stars and dark matter, reinforcing our
initial assumptions \cite{polnarev}.

This seems to contradict the value cited by ZHR of up to 20-40\% of
the central black hole's mass being comprised of dark matter, however
the reason for this discrepancy is clear. ZHR look at the effect of
the diffusion of dark matter and stars into the loss cone via
relaxation, whereas we look at refilling the loss cone by scattering
from massive infalling bodies. As mentioned above, if
the loss cone refilling rate is significant, then all of the dark
matter within 1kpc could be swept up leading to a significant
fraction of a black hole's mass comprising of dark matter. In previous
literature (see e.g. Lightman \& Shapiro 1977) the diffusion rate of
matter into the loss cone was calculated by making three main assumptions:
\\\\
1. The supermassive black holes are fed from shallow cores.\\\\
2. The stellar distribution is spherical or axisymmetric and angular
momentum is conserved.\\\\
3. Star-star relaxation is the dominant process for repopulating the
loss cone.
\\\\
ZHR question these three assumptions in their paper and show that, 
the loss cone refilling rate could be significantly higher than
previously thought. They then go on to add that if there were no
depletion of orbits for the dark matter, that is, if the loss cone
were kept continually filled then some 20-40\% of a supermassive black
hole's mass could be comprised of dark matter. While this is true, it
is somewhat misleading.  

The relaxation time for N particles in a collisionless system and
assuming that each particle moves in the {\it mean} potential of all
the others is given by \cite{binney}: 

\begin{equation}
t_{relax}(r) = \frac{N}{8\ln{N}}\left(\frac{r^3}{GM(r)}\right)^\frac{1}{2}
\end{equation}

Where $M(r)$ is the mass interior to radius $r$, and $N$ is the number
of particles. 

Now, given that the number of dark matter particles must be very very
large, and since $t_{relax} \propto N$, the dark matter-dark matter
relaxation time will be very much longer than the age of the
universe. However, a dark matter particle moving in the granular field
of many stars will behave in the same way a particle of any mass
moving through the same field. As such, the dark matter, even if cold
and with very low self-interaction cross section, can be relaxed in the
same way as the stars {\it by the stars}. 

The question is at what radius does the relaxation time for stars and
dark matter fall below a Hubble time? If we simply model a galaxy as
an exponential disk of stars, a central massive black hole and a power
law dark matter profile, then the relaxation time is given by:

\begin{equation}
t_{relax}(r) =
\frac{N_*}{8\ln{N_*}}\left(\frac{r^3}{G(M_{bh}+M_b(r)+M_{dm}(r)}\right)^\frac{1}{2}
\label{eqn:trelaxzhr}
\end{equation}

Where $M_{bh}$ is the black hole mass, $M_b(r)$ is the mass
distribution of the baryons and $M_{dm}$ is the mass distribution of
the dark matter. We can approximate $N_* \sim M_b(r)/M_\odot$ and so,
for given mass profiles and a given central black hole mass, the relaxation
time can be found as a function of radius. For a Milky Way type
galaxy, the relaxation time is only less than a Hubble time within 2
parsecs of the central black hole. The total matter out to this radius
is not a significant fraction of the central black hole's start
mass. ZHR argue that the relaxation time will be less than given in
equation \ref{eqn:trelaxzhr} due to the relaxation of assumptions 1-3
above. They present some plausible arguments that could reduce the
relaxation time and then state that, should it be sufficiently low that
the loss cone can be kept refilled out to the black hole's sphere of
influence, then 20-40\% of a central supermassive black hole's mass
could comprise of dark matter. It is not clear, however, whether the
relaxation time really could be low enough to produce significant loss
cone refilling. Even if it were, the radius to which the effect can be
considered important must be less than the sphere of influence of the
black hole, and so cannot be important for producing dark matter cores
on the scale of a kiloparsec or so - the kind of core which would be
required to reconcile measurements from rotation curves with cold dark
matter theory. Finally, they do not consider the effect of mergers on
the central matter profile. As stated above, a merging binary black
hole can eject much of the central matter from a galaxy on the scale
of 10-100 parsecs or so. Even if the relaxation time at 100 parsecs
were quite low, it is likely that the matter in this region would be
removed by merging black holes before it would have the chance to be
accreted - either by diffusion due to the coarse grained stellar
potential, as suggested by ZHR, or by scattering from merging
supermassive black holes as suggested here.

\section{Conclusions}
We have shown that supermassive black holes cannot alter a power law
density cusp via accretion, whether during mergers or in the steady
state. This represents another vital piece in the debate surrounding
the nature of dark matter. If winds and bars also prove to be
ineffective at altering a primordial dark matter cusp, we will be
drawn inexorably towards the conclusion that dark matter is more
complex than the simple cold particle envisaged so far.

\section{Acknowledgements}
We would like to thank Mark Wilkinson and the referee for useful
comments and discussion which led to this final manuscript. This work
was funded by a PPARC studentship.

\section{Appendix I - The velocity dispersion in a power law density
profile}
For spherically symmetric, isotropic, systems we have \cite{binney}:

\begin{equation}
\frac{1}{\rho}\frac{d(\rho<v^2>)}{dr} = -\frac{v_c^2}{r}
\label{eqn:p204bt}
\end{equation}

If the density, $\rho(r)$, is given by equation \ref{eqn:power} then
$v_c^2$ is given by equation \ref{eqn:initvel} and equation
\ref{eqn:p204bt} becomes:

\begin{equation}
-\alpha r^{-1}<v^2>+\frac{d<v^2>}{dr} = \frac{-4\pi G
 \rho_0r_0^\alpha}{3-\alpha}r^{1-\alpha}
\end{equation}
$\Rightarrow$
\begin{eqnarray}
<v^2> & = & \frac{2\pi
 G\rho_0r_0^\alpha}{(3-\alpha)(\alpha-1)}r^{2-\alpha} \hspace{0.1in}
 1 < \alpha < 3 \nonumber \\
& = & -2\pi G\rho_0r_0^2x\ln x \hspace{0.4in} \alpha = 1 \nonumber \\
 x & = & \frac{r}{r_0}
\end{eqnarray}

Since the power law density is describing the {\it inner} parts of the
halo, we require $r<r_0$ always and so $x<1$. Thus $\ln x$ will always
be negative such that $<v^2>$ is still positive for $\alpha=1$. Thus
substituting everywhere for $v_c^2$ (see equation \ref{eqn:initvel})
we get:  

\begin{eqnarray}
<v^2> & = & \frac{v_c^2}{2(\alpha-1)} \nonumber \\
& \simeq & v_c^2 \hspace{0.4in} 1 < \alpha < 3 \\
<v^2> & = & -v_c^2\ln x \nonumber \\
& \simeq & v_c^2 \hspace{0.4in} \alpha=1
\end{eqnarray}

\bsp

\label{lastpage}

\end{document}